# Dust Settling in Magnetorotationally-Driven Turbulent Discs I : Numerical Methods and Evidence for a Vigorous Streaming Instability

By

Dinshaw S. Balsara[1] (dbalsara@nd.edu), David A. Tilley[1] (dtilley@nd.edu), Terrence Rettig[1] (trettig@nd.edu), Sean A. Brittain[2] (sbritt@clemson.edu)

**Mailing Address:**

1: Department of Physics
   225 Nieuwland Science Hall
   University of Notre Dame
   Notre Dame, Indiana 46556
   USA

2: Department of Physics
   118 Kinard Laboratory
   Clemson University
   Clemson, South Carolina 29634
   USA

**Phone:** (574) 631-2712

**Fax:** (574) 631-5952

# Dust Settling in Magnetorotationally-Driven Turbulence I : Numerical Methods and Evidence for a Vigorous Streaming Instability

By

Dinshaw S. Balsara[1] (dbalsara@nd.edu), David A. Tilley[1] (dtilley@nd.edu), Terrence Rettig[1] (trettig@nd.edu), Sean A. Brittain[2] (sbritt@clemson.edu)

[1] Department of Physics, University of Notre Dame

[2] Department of Physics, Clemson University

**Abstract**

In this paper we have used the RIEMANN code for computational astrophysics to study the interaction of a realistic distribution of dust grains with gas at specific radial locations in a vertically stratified protostellar accretion disc. The disc was modeled to have the density and temperature of a minimum mass solar nebula and shearing box simulations at radii of 0.3 and 10 au are reported here. The disc was driven to a fully-developed turbulence via the magnetorotational instability (MRI). We find that the inclusion of standard dust to gas ratios does not have any significant effect on the MRI even when the dust sediments to the midplane of the accretion disc. The density distribution of the dust reaches a Gaussian profile regardless of the initial conditions with which the dust was released. The vertical scale heights of these Gaussian profiles are shown to be proportional to the reciprocal of the square root of the dust radius when small spherical dust grains are considered. This result is consistent with theoretical expectation.

The largest two families of dust in one of our simulations show a strong tendency to settle to the midplane of the accretion disc. The large dust tends to organize itself into elongated clumps of high density. The dynamics of these clumps is shown to be consistent with a streaming instability. The streaming instability is seen to be very vigorous and persistent once it forms. Each stream of high density dust displays a reduced RMS velocity dispersion. The velocity directions within the streams are also aligned relative to the mean shear, providing further evidence that we are witnessing a streaming instability. The densest clumpings of large dust are shown to form where the streams intersect.

We have also shown that the mean free path and collision time for dust that participates in the streaming instability is reduced by almost two orders of magnitude relative to the average mean free paths and collision times. The RMS velocities between the grains also needs to fall below a minimum threshold in order for the grains to stick and we show that some of the large dust in our 10 au simulation should have a propensity for grain coalescence. The results of our simulations are likely to be useful for those who model spectral energy distributions (SEDs) of protostellar discs and also for those who model dust coagulation and growth.

**1) Introduction**

The settling of dust in protostellar accretion discs has long been recognized as an important first step in the process of building planets (Safronov 1969, Goldreich & Ward 1973). Several recent observational studies have underscored the importance of dust settling and grain growth in planet formation. Sicilia-Aguilar et al. (2005) have shown that gas-rich discs persist for several million years around low mass stars. Dust mineralogy of the inner discs of Herbig AeBe stars shows a larger fraction of silicate lines in the inner disc, suggesting a possible migration and pile-up of dust relative to the gas (Van Boekel et al. 2004). HST observations of the disc in the HH30 system (Burrows et al. 1996) and theoretical modeling of the spectral energy distribution (SED henceforth) of that system (Cotera et al. 2001) suggests that the dust can grow as well as undergo preferential sedimentation to the disc's mid-plane. Mid-infrared imaging and spectrometry along with millimeter imaging (Natta et al. 2007, Rodmann et al. 2006) suggests the existence of grains up to a cm in size in T Tauri discs. Observations of dust to gas ratios in a range of T Tauri systems observed at various inclinations also indicate that the dust to gas ratio is higher in systems that are seen edge-on (Rettig et al. 2006). Thus observations of dust to gas ratios also suggest that some amount of dust settling to the mid-planes of protostellar accretion discs does indeed take place.

The persistence of gas-rich discs (Sicilia-Aguilar et al. 2005) over timescales of several million years, along with the substantial amounts of gas in the outer planets in our solar system, suggests that planet formation takes place in a gas-rich environment. The core accretion model for the formation of gas giant planets indeed requires the initial formation of a rocky core onto which the gas can accrete (Mizuno 1980, Pollack et al. 1996). Disc instability models for the formation of gas giant planets have also been proposed (Boss 2005) but require cold, massive accretion discs with disc to star mass ratios in excess of 0.01. Such large disc to star mass ratios may be difficult to reconcile with observations, leading us to be mindful of the core accretion model. We therefore see that it is important to be able to understand the settling and growth of dust in accretion discs. We consider dust settling and growth in the next several paragraphs.

The dust in a protostellar disc would settle to the mid-plane in the absence of turbulence. Early studies of dust settling in turbulent accretion discs were carried out by Miyake & Nakagawa (1995) and Dubrulle et al. (1995). Smaller dust grains, with dust sizes smaller than a mean free path in the gas, couple strongly to the gas via an Epstein drag law. The turbulence gives it a stochastic (Brownian-like) motion that allows the small dust grains to resist the pull of gravity towards the disc's mid-plane. Thus the gas and the smaller dust grains remain well-mixed. Larger dust grains couple weakly to the gas via a Stokes drag law, therefore, settling to the disc's mid-plane. Very large dust particles set up their own turbulent wakes that produce an additional drag force relative to the gas. Progressively larger dust couples to the gas with decreasing strength, resulting in smaller scale heights, and therefore greater settling, for larger dust.

The models presented in Miyake & Nakagawa (1995) and Dubrulle et al. (1995) relied on an $\alpha$-viscosity formulation without positing a source for the turbulence. Several hydrodynamical alternatives exist for driving turbulence in protostellar discs. Wiedenschilling (1977, 1980), Ishitsu & Sekiya (2003) and Gomez & Ostriker (2005) suggested that Kelvin-Helmholtz instabilities between a sedimented layer of dust and the overlying gas can drive turbulence in the disc, an idea explored via two-dimensional simulations (Cuzzi et al. 1993) and three-dimensional simulations in Johansen et al. (2006). The dust is also susceptible to preferential radial migration relative to the gas (Nakagawa et al. 1986) and such motions were recognized to be the source of a streaming instability (Goodman & Pindor 2000, Youdin & Goodman 2005). The streaming instability was shown to lead to turbulence when it becomes fully non-linear (Youdin & Johansen 2007, Johansen & Youdin 2007).

While these hydrodynamical instability mechanisms produce a certain level of turbulence in protostellar discs, the strength of the fully saturated turbulence that they produce can also be estimated. Johansen & Youdin (2007) report turbulent Mach numbers that are less than 0.001. We also know that the observed accretion rate on to T Tauri stars requires that an effective $\alpha$ of ~ 0.01 prevails in the disc (Hartmann 1998). The magnetorotational instability (MRI henceforth) (Balbus & Hawley 1991, 1998) is known to produce a level of turbulence that is sufficient for explaining the mass accretion rate in T Tauri systems (Hartmann 1998). It is, however, worth considering Umurhan, Menou & Regev (2007), Umurhan, Regev & Menou (2007), Fromang & Papaloizou (2007) and Fromang et al. (2007) for an emerging viewpoint on the dependence of the MRI on the specifics of the simulation code. Fromang & Papaloizou (2006) and Carballido et al. (2006) have studied the sedimentation of dust in vertically stratified, MRI-unstable discs in the Epstein regime using only 100 particles to represent dust of each size. This would have made it impossible for them to discover some of the collective effects reported in the present paper for larger dust.

In this pair of papers we study the settling of dust in vertically stratified, MRI-unstable discs at a range of radii while generalizing from the Epstein drag approximation. The strategy pursued in this paper is to take the mid-plane density of a minimum mass solar nebula along with the mid-plane temperatures from a radiative transfer model (D'Alessio et al. 1998). At several radial stations in such a nebula, we simulate the MRI to non-linear saturation using a vertically stratified shearing sheet approximation (Goldreich & Lynden-Bell 1965, Goldreich & Tremaine 1978). We then study the settling of dust particles that richly sample space, making it possible for collective instabilities to develop in the simulations. In this work, a distribution of dust particles that represent the Mathis et al. (1977) (MRN henceforth) grain size distribution is used. Cataloguing the scale heights of dust of different sizes, their coupling with the gas and resulting collective instabilities is one of the goals of this pair of papers. The present paper makes an in-depth analysis of two simulations at radii of 0.3 au and 10 au for a minimum mass solar nebula with standard dust to gas ratios. A subsequent paper (Tilley et al., in preparation) will examine these issues for all the radii that we have simulated and also catalogue the effect of gas depletion in the disc. We show that for normal dust to gas ratios the strength of the MRI turbulence is unchanged by the inclusion of dust.

The problem of simulating dust-gas interaction can be carried out by treating the dust as particles as well as a continuum. Prior studies (Johansen & Youdin 2007) have reported some numerical difficulties when the ratio of dust stopping time to dynamical time spanned a large range of numbers in numerical simulations. Because variations on this problem are likely to be of interest in future studies, we also catalogue our numerical methods for treating dust with particulate and continuum representations in this first paper. In the present papers we only treat spherical grains, which constitutes a reasonable initial simplification. Suttner & Yorke (2001) also included fractal dust, which we leave for a subsequent study. Similarly, we work in a single-fluid MHD approximation leaving the study of dust settling in a two-fluid approximation to subsequent papers.

The growth of dust also deserves attention. On the one hand, a robust turbulence in protostellar discs that keeps the smaller, micron-size, dust well-mixed with the gas seems to be essential in ensuring that spectral energy distributions (SEDs henceforth) can be explained (Chiang & Goldreich 1997, D'Alessio et al. 2006). The smaller dust responds to the passive heating from the star, thus keeping the outer layers of the disc well-heated. On the other hand, even modest amounts of relative motion between dust particles can inhibit dust coagulation and growth (Blum & Wurm 2000). The coagulation and growth of dust has been studied quasi-numerically by Dullemond & Dominick (2005) and also via two dimensional numerical simulations by Suttner & Yorke (2001). Based on their simplified models, which represent the dust coagulation rate as a process that is proportional to the square of the number density of the dust, Dullemond & Dominick (2005) find that dust coagulation happens very rapidly. As a result, they had to invoke a fragmentation rate that almost balanced the coagulation rate in order to bring planet formation in line with rates that are observationally supported (Brearley & Jones 1998). One realizes, therefore, that an ability to quantify the relative velocities between dust particles of a given size as a function of height from the disc's mid-plane would go a long way towards enabling better models for coagulation. One of the goals of this pair of papers is to catalogue such information. This paper and its companion are therefore intended to be useful to those who model SEDs and also to those who model dust coagulation and growth.

Having realized that the experimental data of Blum & Wurm (2000) requires the relative velocities between dust particles to be very low in order to achieve dust sticking and growth, several authors began to explore collective processes for dust growth. Sekiya (1998), Youdin & Shu (2002), Garaud & Lin (2004), Garaud et al. (2004) and Weidenschilling (2006) have pointed out that gravitational instabilities are much more likely to be effective in a dust sub-layer that settles to the midplane in situations where the dust to gas ratio in the disc is elevated. The ratio of the dust to gas surface densities need not be fixed at the solar value of 0.01. As pointed out by Youdin & Chiang (2004) and Throop & Bally (2005), photoevaporation of the gas can increase the ratio of the dust to gas surface densities by a factor of ten, making it very favorable to have gravitational instabilities in the dust sub-disc. Alternatively, the inward migration of larger dust grains in a protostellar disc can have the same effect (Nakagawa et al. 1986). Such ideas also serve as the motivation for the suggestion by Goldreich et al. (2004a,b) that a second generation of planet formation might take place via gravitational instabilities in a dust-depleted disc. While dust stopping times may be comparable to or smaller than dynamical times in gas-rich discs, that may not be the case for gas-depleted discs. As a result, having a representation of aerodynamic drag that goes over continuously from the Epstein to the Stokes limit becomes even more important when treating gas-depleted discs. It is one of the further goals of this work to study the settling and velocity dispersion properties of dust particles in gas-depleted discs. As pointed out by Youdin &

Goodman (2005) a gas-depleted disc becomes even more susceptible to collective instabilities such as the streaming instability. An additional goal of this work is to computationally demonstrate that there is a very interesting analogue to the streaming instability in situations where MRI-driven turbulence interacts with dust.

The plan of this paper is as follows. In Section 2 we catalogue our model as well as relevant scales in this problem as a function of radius in a minimum mass solar nebula. In Section 3 we catalogue our methods for treating dust-gas interaction with a semi-implicit and a fully implicit time-integration scheme in the particulate and continuum approximations. In Section 4 we present several results for a small fraction of our simulations. In Section 5 we provide several conclusions. In a subsequent paper (Tilley et al., in preparation) we will present a larger set of runs that sample several radial distances in a minimum mass solar nebula. Sections 2 and 3 are provided for readers who are interested in numerical methodology. Readers who are primarily interested in the astrophysics can skip to Section 4 after quickly reading Section 2.1.

**2) Model set-up, governing equations and relevant scales in the problem**

In the next three subsections we describe our model set-up, catalogue the equations being solved and examine the relevant scales in the problem.

**2.1) Model set-up**

Our simulations physically correspond to a minimum mass solar nebula that is passively heated by a central 0.5 $M_\odot$ star. The disc has a surface mass density given by $\Sigma(r) = 1700 \, (r / 1\text{AU})^{-1.5}$ g cm$^{-3}$ and the run of temperature with radius is taken from D'Alessio et al. (1998). At radii of 0.3 au and 10 au we set up MRI calculations in the shearing sheet approximation, and the relevant simulational parameters are catalogued in Table 1. The temperatures $T_0$ from the models of D'Alessio et al. (1998) are also catalogued in Table 1 as are the sound speeds $c_s$; the simulation at each radial station is taken to be isothermal. Corresponding to the angular frequency $\Omega$ at each radial point of interest we can specify the scale height of the gas $H_{gas} = \sqrt{2}\, c_s/\Omega$ and from it the mid-plane density $\rho_0 = \sqrt{\pi}\, \Sigma(r)/H_{gas}$.

The x, y and z directions in the simulations correspond to the radial, azimuthal and vertical directions in the disc. The computational domain extends over [-$H_{gas}$, $H_{gas}$]X[-$H_{gas}$, $H_{gas}$]X[-3$H_{gas}$, 3$H_{gas}$] in the x, y and z directions and the number of zones used in each of those directions is given in Table 1. The zones are uniform in x and y and ratioed in the z-direction with a ratio of 1.00732 from one zone to the next extending away from the mid-plane. As a result, the smallest zone in the vertical direction has $\Delta z = 0.0144\, H_{gas}$.

The density at each radial station is given by $\rho(z) = \rho_0\, e^{-z^2/H_{gas}^2}$ and the magnetic field is produced by a vector potential that goes as

$$A_x = \begin{cases} \left(32\rho_0 c_s^2/\pi\beta_0\right)^{1/2} H_{gas} \cos(2\pi x/H_{gas}) \cos(2\pi y/H_{gas}) & |z| \leq 1.25 H_{gas} \\ \left(32\rho_0 c_s^2/\pi\beta_0\right)^{1/2} H_{gas} \cos(2\pi x/H_{gas}) \cos(2\pi y/H_{gas}) e^{-[(|z|-1.25 H_{gas})/H_{gas}]^4} & 1.25 H_{gas} < |z| < 2.5 H_{gas} \\ 0 & |z| \geq 2.5 H_{gas} \end{cases}$$

where $\mathbf{B} = \nabla \times \mathbf{A}$. The models for gas density and magnetic field presented here are closest to those in Miller & Stone (2000).

When initializing particles we use 5 logarithmically spaced grain sizes ranging from 0.25 μm to 2.5 mm. We use 1,859,674 particles per family of grains. The particles were initialized over [-2.5$H_{gas}$, 2.5$H_{gas}$] in the z-direction. All particles corresponding to a given grain size had the same mass. They were also initialized with the gas scale height and were given the same Gaussian distribution in the z-direction as the gas. The particles follow an MRN grain distribution with the number density varying with grain size "a" as $N(a) \sim a^{-3.5}$. While Bouwman et al. (2003) have measured the grain distribution in HD100546 and found a flatter distribution $N(a) \sim a^{-2}$, we use an MRN distribution here simply because it is the more traditional choice. As future observations provide more guidance on grain sizes, a thorough study of the influence of grain growth, it might be worth undertaking a more thorough study that incorporates the role of grain evolution. The midplane density of all the dust species added together was taken to be one-hundreth the gas density $\rho_0$.

**2.2) Governing equations**

We utilize the equations of ideal, isothermal magnetohydrodynamics to evolve the gas in the disc. By doing so, we implicitly assume that the gas is well-ionized, so that it can couple strongly to the magnetic field. This assumption can, however, become problematic in the presence of a dead zone (Gammie 1996). Recent results have suggested that the cosmic-ray energy density may be enhanced by several orders of magnitude (Fatuzzo, Adams & Melia 2006), and hence the role of dead zones may be reduced. Furthermore, much of the emission in the SEDs of protostellar discs comes from beyond 5 AU, exterior to the region where dead zones can form.

The equations of ideal MHD can be found in many sources, and the strategies we use for solving them can be found in Balsara 1998a,b, Balsara & Spicer 1999a,b, and Balsara 2004. To the momentum equation we add an additional force term that represents the momentum transfer between the dust and the gas:

$$\frac{\partial}{\partial t}(\rho_g v_g) = f_g - \sum_j \alpha_j \rho_j \rho_g (v_g - v_j) \qquad (1)$$

where $f_g$ is the sum of the rest of the forces on the gas, $\alpha_j$ is a frictional coupling constant between the dust and gas, and the sum is over all the species of dust "j".

We present two methods for the treatment of the dust, which we assume is neutral and will not thus interact directly with the magnetic field. In general, dust grains are likely to be charged; to treat their motion in a magnetized plasma, however, would require a more sophisticated equation of motion, such as a guiding centre approximation (e.g. Yan & Lazarian 2003). Furthermore, the charge per grain is ambiguous and would require a detailed charging model that is outside the scope of this work. The first method evolves the dust as an Eulerian fluid on the same mesh on which we evolve the gas. The other method uses Lagrangian superparticles to track the evolution of the dust. For the Eulerian method, the friction term that is added to the momentum equation for each dust family "j" is the same as the one in Eq. 1, but with opposite sign to conserve momentum:

$$\frac{\partial}{\partial t}(\rho_j \mathbf{v}_j) = \mathbf{f}_j - \alpha_j \rho_j \rho_g (\mathbf{v}_j - \mathbf{v}_g) \tag{2}$$

($\mathbf{f_j}$ is the sum of the rest of the forces acting on the dust). For the particle treatment of the dust, the equation of motion for an individual superparticle, which we label "k", belonging to dust family "j" is

$$\frac{d\mathbf{v}_k}{dt} = \mathbf{a}_k - \alpha_j \rho_g (\mathbf{v}_k - \mathbf{v}_g)$$

(3)**Error! Bookmark not defined.**

Here, $\mathbf{a_k}$ is the acceleration due to the other forces acting on the particle; we have divided out the particle mass which figures into Eq. (**Error! Bookmark not defined.**3).

Eq. (1) must be solved simultaneously with either Eq. (2) or (3). To get the desired second-order accuracy, we utilize a predictor-corrector formulation. Our strategy is comprised of a fully implicit predictor stage that gives us stability, coupled with a half-implicit corrector stage that gives us the second-order accuracy. One challenge in creating an implicit strategy for Eqs. (1)-(3) is that the friction coefficient $\alpha_j$ is itself velocity-dependent, and requires further discussion.

The form of the friction constant $\alpha_j$ will depend on whether the friction forces are in the Epstein or Stokes regimes. The Epstein drag law is appropriate when the size of the particles is smaller than the mean free path of the gas; the Stokes drag law is used when the particle size is larger than the gas mean free path. We use the formulation in Yorke (1979) that provides a transition for the drag coefficient between the subsonic and supersonic drag regimes. The inclusion of this relative velocity term gives us the potential for transonic flow; while the turbulence in the disc never becomes supersonic, we do find that our simulations develop velocities of ~0.5 $c_s$. The Epstein regime drag coefficient is

$$\alpha_{j, \text{Epstein}} = \frac{4\sigma_j}{3m_j} \sqrt{c_s^2 + |\mathbf{v}_j - \mathbf{v}_g|^2} \tag{4}$$

For spherical dust grains, the mass of a dust grain is $m_j = \frac{4}{3}\pi\rho_{grain} a_j^3$ and cross-section of a grain is $\sigma_j = \pi a_j^2$, where $\rho_{grain}$ is the density of the material that comprises the dust and $a_j$ is the radius of the dust grain. The Stokes regime occurs when the particle sizes become comparable to the mean free path. The Stokes regime drag coefficient we use is (Weidenschilling 1977; Cuzzi et al. 1993)

$$\alpha_{j,\,Stokes} = \frac{3D}{8 a_j \rho_{grain}} |v_j - v_g|$$

$$D = \begin{cases} 24/\text{Re}_{grain} & \text{if } \text{Re}_{grain} \leq 1 \\ 24/\text{Re}_{grain}^{0.6} & \text{if } 1 < \text{Re}_{grain} \leq 800 \\ 0.44 & \text{if } \text{Re}_{grain} > 800 \end{cases} \quad (5)$$

$$\text{Re}_{grain} = 2 a_j |v_j - v_g|/\upsilon$$

where $\upsilon$ is the molecular viscosity of the gas. The transition between these two regimes occurs at $a_j = \frac{9}{4}\lambda_{MFP}$ (Cuzzi et al. 1993), where $\lambda_{MFP}$ is the mean free path in the gas.

The velocity terms in the prescriptions for the Epstein and Stokes drag laws (Eqs. (4) and (5)) require some sensitivity in how they are incorporated into the time-stepping strategy. In practice, the velocity-dependence will start to have an effect only when velocities become transonic, or when the grain sizes become very large. As a result, we can form simple approximations to $\alpha_j$ without seriously harming the quality of our solutions.

**II.c) Relevant scales in the problem**

A dust grain that is moving relative to a uniform region of gas would lose its relative motion in a characteristic stopping time. The stopping time for dust-gas friction in Eqs. (1-4) is defined by $\tau_s = 1/\alpha_j \rho_g$. The ratio of the friction time scale to the orbital time scale is thus $\tau_s/\tau_{orbit} = \Omega/(2\pi\alpha_j \rho_g)$. Using the Epstein drag law in the subsonic regime, and utilizing the scalings of the gas disc given in Section 2.1 and a grain material density of 2 g cm$^{-3}$, this ratio becomes

$$\frac{\tau_s}{\tau_{\text{orbit}}} = \sqrt{\frac{1}{2\pi^3}} \frac{\rho_{\text{grain}} a_j}{\Sigma(r)} \approx \left(1.493 \times 10^{-4}\right) \left(\frac{a_j}{1 \text{ cm}}\right) \left(\frac{r}{1 \text{ AU}}\right)^{1.5} \quad (6)$$

For centimeter-sized grains at 10 AU, this ratio is 0.03. The time taken by a dust particle that is released at one scale height above the midplane of the disc to reach the midplane of the disc is given by the free fall time $\tau_{\text{fall}} = \frac{\pi}{2} \Omega^{-1}$ in the absence of retarding frictional forces.

### III) Treating dust-gas interaction

The coupling via friction of the dust and gas can have a very large effect on the evolution of the dust, as can be seen by inspection of Eqs. (2) & (3). Furthermore, the stopping time for the dust of family "j", $\tau_s = 1/\alpha_j \rho_g$, will vary by a factor of $e^4 \approx 55$ over two scale heights in the stratified disc for dust of a particular grain size. It will similarly vary between grains of different sizes due to both the different friction coefficients $\alpha_j$ and the MRN size distribution. As a result, an implicit method is desirable in order to prevent small friction times from limiting the time step, and to guarantee stability. We formulate below an implicit-explicit scheme below that is second-order accurate in space and time that allows us to avoid the restrictive explicit time step.

We divide the time update into a predictor step that advances the fluid variables from the time at the beginning of the time step, $t^i$, to an intermediate time $t' = t^i + \Delta t/2$, and a corrector step that updates the fluid variables to the next full time step $t^{i+1} = t^i + \Delta t$. Some parts of the formulation presented here parallels the formulation for two-fluid MHD in Tilley & Balsara (2008).

### 3.1) The predictor step for dust treated as continuum

We can write the predictor step for the dust of family "j" as a fully implicit update:

$$\rho'_j \mathbf{v}'_j - \rho^i_j \mathbf{v}^i_j = \frac{\Delta t}{2} \mathbf{f}_j - \alpha_j \rho'_j \rho'_g \frac{\Delta t}{2} \left(\mathbf{v}'_j - \mathbf{v}'_g\right) \quad (7)$$

and simultaneously the fully implicit predictor step for the gas as

$$\rho'_g v'_g - \rho^i_g v^i_g = \frac{\Delta t}{2} f_g - \sum_j \alpha_j \rho'_j \rho'_g \frac{\Delta t}{2}(v'_g - v'_j) \quad (8)$$

Here we denote variables at time step i as unprimed and variables at the time step i+1/2 with a prime. We can express this in a matrix form as

$$\begin{bmatrix} \rho'_1 + \alpha_1 \rho'_1 \rho'_g \frac{\Delta t}{2} & \cdots & 0 & -\alpha_1 \rho'_1 \rho'_g \frac{\Delta t}{2} \\ \vdots & \ddots & \vdots & \vdots \\ 0 & \cdots & \rho'_j + \alpha_j \rho'_j \rho'_g \frac{\Delta t}{2} & -\alpha_j \rho'_j \rho'_g \frac{\Delta t}{2} \\ -\alpha_1 \rho'_1 \rho'_g \frac{\Delta t}{2} & \cdots & \alpha_j \rho'_j \rho'_g \frac{\Delta t}{2} & \rho'_g + \left(\sum_j \alpha_j \rho'_j\right)\rho'_g \frac{\Delta t}{2} \end{bmatrix} \begin{bmatrix} v'_1 \\ \vdots \\ v'_j \\ v'_g \end{bmatrix} = \begin{bmatrix} \rho^i_1 v^i_1 + \frac{\Delta t}{2} f_1 \\ \vdots \\ \rho^i_j v^i_j + \frac{\Delta t}{2} f_j \\ \rho^i_g v^i_g + \frac{\Delta t}{2} f_g \end{bmatrix}$$

(9)

The matrix on the left-hand side of Eq. (9) can be easily inverted analytically in a symbolic mathematics program to solve for $v'_j$ and $v'_g$. In doing so, we calculate the densities of the gas and dust at the predicted time step ($\rho'_g$ and $\rho'_j$) in an explicit manner.

**3.2) The predictor step for dust treated as particles**

The predictor step for a particle treatment of the dust proceeds in a similar manner as the continuum scheme. However, it would be excessively complicated to formulate a purely implicit strategy in both the dust and the gas, as the superparticles for the dust of different grain sizes will not be collocated with each other in general. We can rely on the fact that the dust mass will generally always be much less than the mass in gas, so that explicit time steps are feasible for the gas while the dust update remains implicit.

We use triangular-shaped cell (TSC) interpolation to interpolate between the grid zones and the particle locations. The equation for the update of a dust superparticle "k" of grain size "j" is

$$v'_k - v^i_k = \frac{\Delta t}{2} a^i_k - \frac{\Delta t}{2} \alpha_j \rho^i_g (v'_k - v^i_g) \quad (10)$$

where here $\rho_g$ and $v_g$ are the gas density and velocity interpolated to the location of the particle k. Eq. (10) can be solved for $v'_k$

$$\mathbf{v'}_k = \left[\mathbf{v}_k^i + \frac{\Delta t}{2}\mathbf{a}_k^i + \alpha_j \rho_g^i \frac{\Delta t}{2}\mathbf{v}_g^i\right]\left[1 + \alpha_j \rho_g^i \frac{\Delta t}{2}\right]^{-1} \tag{11}$$

With $\mathbf{v'}_k$ obtained the update for the gas can be performed:

$$\rho'_g \mathbf{v'}_g = \rho_g^i \mathbf{v}_g^i + \frac{\Delta t}{2}\left[\mathbf{f}_g^i - \sum_k W_{TSC}(\mathbf{x} - \mathbf{x}_k^i)\alpha_j m_k \rho_g^i (\mathbf{v}_g^i - \mathbf{v'}_k)\right] \tag{12}$$

Here $\mathbf{x}$ is the position of the zone centre, $\mathbf{x}_k$ is the position of particle k, and $m_k$ is the mass of superparticle k. $W_{TSC}$ is the TSC weighting function.

### 3.3) The corrector step for dust treated as continuum

We build the half-implicit corrector step by averaging the friction force between the dust and the gas at the initial time step "i" and the next time step "i+1":

$$\rho_j^{i+1}\mathbf{v}_j^{i+1} - \rho_j^i \mathbf{v}_j^i = \Delta t \mathbf{f'}_j - \alpha_j \rho_j^{i+1}\rho_g^{i+1}\frac{\Delta t}{2}(\mathbf{v}_j^{i+1} - \mathbf{v}_g^{i+1}) - \alpha_j \rho_j^i \rho_g^i \frac{\Delta t}{2}(\mathbf{v}_j^i - \mathbf{v}_g^i) \tag{13}$$

$$\rho_g^{i+1}\mathbf{v}_g^{i+1} - \rho_g^i \mathbf{v}_g^i = \Delta t \mathbf{f'}_g - \sum_j \alpha_j \rho_j^{i+1}\rho_g^{i+1}\frac{\Delta t}{2}(\mathbf{v}_g^{i+1} - \mathbf{v}_j^{i+1}) - \sum_j \alpha_j \rho_j^i \rho_g^i \frac{\Delta t}{2}(\mathbf{v}_g^i - \mathbf{v}_j^i) \tag{14}$$

We can again write this as a matrix equation that can be inverted analytically.

$$\begin{bmatrix} \rho'_1 + \alpha_1 \rho'_1 \rho'_g \frac{\Delta t}{2} & \cdots & 0 & -\alpha_1 \rho'_1 \rho'_g \frac{\Delta t}{2} \\ \vdots & \ddots & \vdots & \vdots \\ 0 & \cdots & \rho'_j + \alpha_j \rho'_j \rho'_g \frac{\Delta t}{2} & -\alpha_j \rho'_j \rho'_g \frac{\Delta t}{2} \\ -\alpha_1 \rho'_1 \rho'_g \frac{\Delta t}{2} & \cdots & \alpha_j \rho'_j \rho'_g \frac{\Delta t}{2} & \rho'_g + \left(\sum_j \alpha_j \rho'_j\right)\rho'_g \frac{\Delta t}{2} \end{bmatrix} \begin{bmatrix} \mathbf{v'}_1 \\ \vdots \\ \mathbf{v'}_j \\ \mathbf{v'}_g \end{bmatrix} = \begin{bmatrix} \rho_1^i \mathbf{v}_1^i + \Delta t \mathbf{f'}_1 - \alpha_1 \rho_1^i \rho_g^i \frac{\Delta t}{2}(\mathbf{v}_1^i - \mathbf{v}_g^i) \\ \vdots \\ \rho_j^i \mathbf{v}_j^i + \Delta t \mathbf{f'}_j - \alpha_j \rho_j^i \rho_g^i \frac{\Delta t}{2}(\mathbf{v}_j^i - \mathbf{v}_g^i) \\ \rho_g^i \mathbf{v}_g^i + \Delta t \mathbf{f'}_g - \sum_j \alpha_j \rho_j^i \rho_g^i \frac{\Delta t}{2}(\mathbf{v}_g^i - \mathbf{v}_j^i) \end{bmatrix} \tag{15}$$

The matrix on the left-hand side is the same as the one for the predictor step. The drag coefficients $\alpha_j$ and the non-friction forces $f'_g$ and $f'_j$ are calculated at the time-centered time step i+1/2, calculated in the predictor step.

**3.4) The corrector step for dust treated as particles**

To form the half-implicit corrector step for the particle treatment of the dust, we calculate the friction coefficient at the intermediate time step calculated in the predictor time step.

$$v_k^{i+1} - v_k^i = \Delta t a_k^i - \frac{\Delta t}{2}\alpha'_j \rho'_g \left(v_k^{i+1} - v_g^{i+1}\right) - \frac{\Delta t}{2}\alpha'_j \rho'_g \left(v_k^i - v_g^i\right) \tag{16}$$

We will not be able to simultaneously calculate $\mathbf{v}_g^{i+1}$ implicitly along with $\mathbf{v}_k^{i+1}$, but we can approximate this term in Eq. 16 by calculating its update in the absence of dust-gas friction. This approximation will be valid as long as the density of dust in a particular family is less than the gas density. The update for $\mathbf{v}_k^{i+1}$ then becomes

$$v_k^{i+1} = \left[v_k^i + \Delta t a_k^i + \alpha'_j \rho'_g \frac{\Delta t}{2} v_g^{i+1} - \alpha'_j \rho'_g \frac{\Delta t}{2}\left(v_k^i - v_g^i\right)\right]\left[1 + \alpha'_j \rho'_g \frac{\Delta t}{2}\right]^{-1} \tag{17}$$

We can then calculate the correction to the gas velocity by averaging the contributions to the gas momentum at the beginning and end of the time step:

$$\rho_g^{i+1}\mathbf{v}_g^{i+1} = \rho_g \mathbf{v}_g - \frac{\Delta t}{2} \sum_k \left[W_{\text{TSC}}(x_{\text{grid}} - x_k^{i+1})\left(\mathbf{v}_g^{i+1} - \mathbf{v}_k^{i+1}\right) - W_{\text{TSC}}\left(x_{\text{grid}} - x_k\right)\left(\mathbf{v}_g - \mathbf{v}_k\right)\right] m_k \alpha'_k \rho'_g$$
$$\tag{18}$$

**3.5) Test of the streaming instability**

Youdin & Goodman (2005) showed that an unstratified disc that has dust and gas moving at different velocities could experience an instability that is analogous to the two-stream instability in plasma physics. Johansen & Youdin (2007) and Youdin & Johansen (2007, YJ07 henceforth) further developed this idea and presented the eigenvectors of the instability for particular dust-to-gas ratios and frictional dust-gas coupling constants. We utilize those eigenvectors to test whether our semi-implicit method can reproduce the results of YJ07. Our goal is to demonstrate that our code is capable of capturing such collective instabilities should the occur in the problem.

For this test problem, we were required to alter our setup slightly from that described in Section 2.1. In particular, we remove the gravitational force and the density stratification in the vertical direction, so that the initial density is constant everywhere in the computational domain. Furthermore, the test problem described by YJ07 has a net drift between gas and dust incorporated into it that arises from partial support due to the pressure gradient in the gas, leading to sub-Keplerian rotation in the gas component of the disc. The coordinate system used in YJ07 differs slightly from the shearing box formulation that we use as the Keplerian shear is removed from the velocities; this leads to an additional inertial force that must be added to the momentum and continuity equations. The eigenvectors in YJ07 are presented for a two-dimensional system in 'x' and 'z' (corresponding to the radial and vertical coordinates; azimuthal symmetry is assumed). Finally, YJ07 use a single species of dust that is characterized by a constant stopping time $\tau_s = 0.1\Omega^{-1}$, which in our nomenclature is equivalent to $\alpha\rho_g = \tau_s^{-1}$.

We initialize the eigensystem described by Equations 7 and 10 of YJ07 for the two examples presented there – one with a dust-to-gas ratio of 3 ('linA' in the nomenclature of YJ07) and one with a dust-to-gas ratio of 0.2 ('linB' in their nomenclature). We display the growth rates from the resulting simulations as a function of grid resolution in Figure 1, showing both the growth rates from simulations where the dust is treated as a continuum fluid (solid line), and dust treated as particles with an average of 16 particles per zone.

We see that we do fairly well reproducing the growth rates for the 'linA' model. We get rapid convergence to the predicted eigenvalue in each of the quantities except the gas density and transverse velocity. We get slower convergence in the 'linB' model, and see that the growth rate in the densities of both gas and dust convergence to a value about 10% larger than that predicted by Youdin & Johansen (2007).

**4) Results**

In Section 4.1 we catalogue the turbulence generated in the disc by the MRI instability. It has often been conjectured that Kelvin-Helmoltz instabilities between a thin sheet of dust at the mid-plane of a stratified disc can also be a significant source of turbulence. Since our simulations do produce a thin layer of large dust grains in the disc's mid-plane, in Section 4.1 we evaluate the importance of the Kelvin-Helmoltz instability relative to the MRI. Section 4.2 examines dust to gas scale heights in a few disc models for a range of dust sizes. Section 4.3 catalogues the RMS velocity dispersion in the dust as a function of height away from the disc's midplane. In Section 4.4 we provide evidence for a vigorous streaming instability that arises in the large dust grains that sediment to the disc's midplane. In Section 4.5 we evaluate dust mean free paths, stopping times and the r.m.s. velocities between dust grains for the large dust that is undergoing streaming instability. These results are used to draw inferences for grain collisions, coagulation and growth.

**4.1) Cataloguing the magnetorotational instability and the back-reaction of dust on MRI-driven turbulence**

Our simulations were run without dust until the MRI had reached a steady state ( 6 orbits at 0.3 au, 7.5 orbits at 10 au). We then added the dust particles as described in Section 2.1. The evolution of the magnetic energy (thick solid line), kinetic energy (thin solid line), and total (kinetic + magnetic) energy (dashed line) is shown in Figure 2a for the 10 au simulation. For the same 10 au simulation we plot the evolution of the magnetic stress $\alpha_{\mathrm{mag}} = -\int d^3x \, B_x B_y / 4\pi P_0$ (thick solid line; $P_0$ is the gas pressure in the midplane) and kinetic stress $\alpha_{\mathrm{kin}} = \int d^3x \, \rho v_x v_y / P_0$ (thin solid line) in Figure 2b, along with the total stress (dashed line). The kinetic stress term is quite noisy, but the stress terms indicate the MRI turbulence has reached a steady state. We find that our initial conditions lead to slightly smaller values for the magnetic, kinetic and total stresses than that found by Fromang & Papaloizou (2006) but that is not inconsistent with the higher resolutions used here. Fromang & Papaloizou (2007) and Fromang et al. (2007) have indeed shown that there is a resolution-dependence in simulations of the MRI when non-ideal terms are not included, as was the case for the simulations reported here. The level of the turbulent stresses that we get in our simulations are, nevertheless, consistent with the values needed for sustaining accretion in T Tauri discs (Hartmann 1998).

Weidenschilling (1977,1980) has claimed that a vigorous Kelvin Helmholtz instability between a sedimented layer of dust in the midplane of the disc and the rest of the disc gas could be a source of disc turbulence. We do not examine this phenomenon for all possible discs and all possible mass loading by the dust as was done in Cuzzi et al. (1993). However, the larger dust in our 10 au simulation does sediment to a thin layer in the midplane of our stratified disc as will be shown in the next subsection. It therefore becomes interesting to examine the role of any Kelvin-Helmholtz instability in enhancing the MRI-driven turbulence in our simulations. For that reason we ran one version of our 10 au simulation without dust well past 7.5 orbital times. This dust-free simulation was then compared to the analogous simulation in which the dust had been introduced at 7.5 orbital times and the difference between the integrated values from the dust-free and dust-laden simulations were plotted out as a function of time. Figure 2c shows the difference in the evolution of the magnetic, kinetic and total energies. Figure 2d similarly shows the difference between the magnetic, kinetic and total stresses. The legends in Figures 2c and 2d parallel those in Figures 2a and 2b respectively. Figure 2c shows us that there is no systematic change in the magnetic energy. We do see that the kinetic energy of the dust-laden simulation decreases by about 3% . This is consistent with the fact that the dust, especially as it sediments to the disc's midplane, provides an extra mass-loading to the turbulence. Since the magnetic energy exceeds the kinetic energy by a substantial margin in Figure 2a, the total energy shows very little change when dust is included. Similarly, the total stress in Figure 2d shows little change due to the inclusion of the dust in the MRI turbulence.

**4.2) Dust to gas scale heights**

As mentioned in Section 2.1, the simulations are initialized by releasing all dust families within 2.5 times the gas scale height in the turbulent accretion discs. All families of dust then oscillate about the disc's midplane while they seek out their own scale heights consistent with the driving from the MRI-turbulence. We therefore evaluate the ratio of dust to gas scale heights and plot them as a function of time in Figures 3a and 3b for the 0.3 au and 10 au simulations respectively. The time in Figures 3a and 3b corresponds to the time, measured in units of an orbital period, after the first release of the dust. We see from these figures that the dust scale heights reach a steady state within a few orbits. While the largest dust in Figure 3a only has a scale height that is about half that of the gas, Figure 3b shows us that the larger dust grains at 10 au undergo a very significant rearrangement.

We plot the mean density of dust as a function of height above and below the midplane in Figure 4, averaged over the entire domain in x and y, at two stations within the disc: 0.3 au (Figure 4a) and 10 au (Figure 4b). Because the turbulence is sub-sonic, the gas density tracks its initial Gaussian profile extremely well and is, therefore, not shown. The different symbols show the dust density for each of the five dust families we use. Dashed lines that pass through each of the five different symbols are fits to the dust distribution using a Gaussian profile. We see that the mean density profiles for all of the dust species are approximated very nicely by the Gaussian profiles. The only deviation from the Gaussian comes in the extreme wings of the dust profiles, which are strongly depleted in dust. The simulation at 10 au becomes stratified to a significantly greater degree, as the 2.5 mm dust contracts into a very thin layer while the smaller dust grains are progressively better mixed with the gaseous disc. This thin layer is nevertheless captured with about 6 zones on either side of the disc's midplane, giving us confidence that the scale height that we extract for the largest dust at 10 au is reasonably accurate. Our results bear out the claim from Dubrulle, Morfill & Sterzik (1995) that the dust equilibrates to a Gaussian profile even if it undergoes a significant structural rearrangement in doing so. For the disc radii examined here, this equilibration process only takes a few orbits. The astrophysical significance of this result is that even if dust-rich material rains down on an astrophysical disc or even if the disc structure undergoes a significant rearrangement, the dust distribution always reaches a simple steady state and that distribution is always described by a simple Gaussian profile. It takes several stopping times (Eq. 6) to establish a Gaussian dust distribution.

We plot the scale heights of the dust of each grain size as a function of grain size in Figure 5. In both simulations, we see that the larger grain sizes have a smaller scale height relative to the gas. The decrease in scale height is much steeper at 10 au, and approximates a power-law of the form $H_{dust}/H_{gas} \propto a^{-1/2}$. Sedimentation-diffusion models (Dubrulle, Morfill & Sterzik 1995; Fromang & Papaloizou 2006) predict that the dust scale height in the steady-state balance between gravitational sedimentation and turbulent diffusion varies with dust stopping time as $H_{dust}/H_{gas} \propto (\tau_s/\tau_{orbit})^{-1/2}$. For the Epstein drag law (Eq. 4), this leads to a relationship between dust scale height and size of $H_{dust}/H_{gas} \propto a^{-1/2}$ (see Eq. 6). Thus, the power-law behaviour we see at large dust sizes at 10 au reinforces this idea that the dust scale height can be characterized by sedimentation and turbulent diffusion processes. The departure from the relation $H_{dust}/H_{gas} \propto a^{-1/2}$ seen at small dust sizes at 10 au, and all dust sizes at 0.3 au, results from the limitation that the dust scale height cannot be larger than the gas scale height. We note that the Stokes drag law (Equation 5) for Re < 1 leads to $H_{dust}/H_{gas} \propto a^{-1}$. For reference, we plot the slopes corresponding to power laws of -1/2 and -1 in Figure 5.

**4.3) Velocity dispersion in the dust**

Figure 6 plots the total velocity dispersion (the shear in the disc has been removed from the y-velocity dispersion) as a function of height above and below the midplane. It is apparent that at both 0.3 au (Figure 6a) and at 10 au (Figure 6b), that the dust velocity dispersion is largely independent of the height above the disc. The magnitude of the velocity dispersion is the same for all of the dust species. The close match between the fluctuations in the velocity dispersions as a function of height is what we would expect if the dust were closely coupled to the gas. From Figure 2a we see that the RMS velocity of the gas, $v_{rms}$, is related to the sound speed, $c_s$, by $v_{rms} = 0.06 c_s$. By defining an eddy turn over time for the disc's turbulence by $\tau_{eddy} = H_{gas}/v_{rms}$ we realize that the eddy turn over time is related to the orbital time as $\tau_{eddy} = 3.75\, \tau_{orbit}$. Eqn. (6) then tells us that even for the largest 2.5 mm grains in our simulation at 10 au we have $\tau_s/\tau_{eddy} = 1.2 \times 10^{-3}$ thus ensuring that the grains are very effectively stirred by the turbulence.

**4.4) A novel streaming instability in MRI-driven turbulence interacting with dust**

Figure 7 shows the logarithm (base 10) of the gas and dust densities in four planes in the 10 au simulation: the x-y plane at z=0, the x-y plane at z = 3 $H_{gas}$, the x-z plane at y = $H_{gas}$, and the y-z plane at x = $H_{gas}$ (In other words, if we envision the computational domain as being a square prism then we are removing the interior of the computational domain and showing the furthest three faces. As most of the interesting dust activity occurs in the disc's midplane, we show that plane as well). The upper two frames show the gas density and the density of the 0.25 µm dust. The middle two frames show the 2.5 µm dust density and the 25 µm dust density. The lowest two frames show the 250 µm dust density and the 2.5 mm dust density. The low levels of turbulence in the disc lead to a very smooth appearance in the gaseous disc. The density fluctuations in the midplane of the 0.25 µm dust are primarily on small scales, and range over half an order of magnitude in density. We see that at 10 au the 25 µm dust has experienced significant settling towards the midplane. The 2.5 mm dust has collapsed into a very thin sheet. It is also apparent that the large 2.5 mm dust has collected into a few large concentrations in the midplane, while other regions in the midplane are highly evacuated. Since such clumping could constitute a first step towards planetismal formation, we study the large dust at 10 au in detail in this subsection. Once the 2.5 mm and 250 µm dust in the 10 au simulation settles to the disc midplane, we find that such clumps keep forming repeatedly in the simulation. The clumps in the 2.5 mm and 250 µm dust families in Figure 7 overlie each other in the disc midplane giving us our first clue that we are witnessing a self-organizing phenomenon. These clumps are indeed a very persistent feature in the simulation and some amount of clump-forming activity can always be seen in the simulation. Because the simulations did not have self-gravity, the clumps are sheared away by the overall shear in the simulations. It is, however, conceivable that with the inclusion of self-gravity and grain growth, the largest such clumps could become self-gravitating. The 0.3 au simulation shows a much smoother distribution for all families of dust. Since this would be anticipated from Figure 4, we do not show the gas and dust at 0.3 au.

To further understand what is happening to the large dust in the midplane, we plot in Figure 8 histograms of the dust density of each family of dust in the two layers of the grid that comprise the midplane of our simulations (i.e. the two layers of zones that border z=0; this is equivalent to the volume between $z = -0.014\ H_{gas}$ and $z = 0.014\ H_{gas}$.) To enable the reader to make comparisons, we show the density histograms of the dust families at 0.3 au (Figure 8a) and at 10 au (Figure 8b). As all of the zones that go into the histogram are at the same distance from the midplane, none of the density variation in Figure 8 comes from the density stratification in the disc; it is purely due to the fluid flows forming in the midplane as well as the natural tendencies for self-organization in the dust. We see in Figure 8a that at 0.3 au, the shape of the distribution of dust density in the midplane zones is the same for all five dust families, with the position of the peak shifting to lower densities for larger grains due to the lower abundances of those grains from the MRN grain size distribution. In Figure 8b the three smallest grain families have similar density histograms. However for the two largest grain families we see that about a quarter of the volume in the midplane has been significantly depleted (densities more than two orders of magnitude less than the mean density of the dust in that family). These are the evacuated regions we saw in Figure 7. We also see that the maximum densities for these large grains are even larger than the maximum densities in the smaller grains. Sedimentation accounts for a significant fraction of this, but sedimentation is not the sole reason for the high densities in the large dust. The ratio of the maximum dust density of 2.5 mm dust grains in the midplane layer to the mean density in the midplane layer at 10 au is 57.9, and for the 250 µm dust the same ratio is 30.7. The ratios of maximum density to mean density in the midplane for the three smallest dust families in the same simulation are consistently between 5.7 and 6.0. Thus, we see strong evidence for clumping of large dust grains once they become highly stratified in the disc. We proceed to provide further evidence that this is due to the streaming instability of Youdin & Johansen (2007).

The streaming instability is a natural consequence of collections of solid objects moving at a coherent velocity with respect to a fluid. Real world analogies to the physics of this instability can be seen in a flock of birds flying in formation, or a peloton of cyclists moving together. In such situations, the leading objects cause the air to move with them, thus reducing the wind resistance for the objects that follow in the formation. Youdin & Goodman (2005) showed that dust grains moving through a sub-Keplerian gaseous disc would tend to organize themselves similarly. I.e. it is energetically favorable for the dust grains to be clumped up. A central attribute of such an instability consists of the grains in the high density clumps having a low velocity dispersion amongst them. In Figure 9 we show a scatter plot for the RMS velocity dispersion for the dust in each midplane zone versus the density of dust in that zone. The mean shear was removed from the velocity dispersion in Figure 9. Figure 9a shows this information for the largest grains (2.5 mm) in the midplane of the 0.3 au simulation while Figure 9b shows the same information for the same grains at 10 au. We display all the particles that are within one scale height of the midplane for the largest dust. We see from Figure 9b that the 2.5 mm grains at 10 au show several dense clumps with densities that are several times greater than the mean density. Furthermore, these ultra-dense regions have the lowest velocity dispersions. Figure 9a shows very few ultra-high dense clumps though it does show a smaller trend for having low velocity dispersions in denser regions. The 250 μm dust at 10 au shows a similar systematic trend which gives us assurance that the phenomenon that we are witnessing is not a consequence of the very small scale height of the 2.5 mm dust at 10 au.

While Figure 9 showed that denser clumps in the streaming instability have smaller RMS velocity dispersions, a robust streaming instability also requires all the dust particles to be moving in the same direction. It is therefore worthwhile to consider the directions of the velocities. Figure 10a shows a color plot of the logarithm (base 10) of the midplane density that was shown in Figure 7d for the largest grains. If these density clumps are moving with the same velocity then the grains that make up those clumps should be moving in the same direction. Figure 10b shows the same information but this time we plot the individual particle positions. The particles have been colour coded according to the direction of their velocity and the colored pie chart at the bottom of Figure 10b shows the colour coding used for the direction of the velocity in the xy-plane. The mean shear has been removed. We see that the particles have arranged themselves along high density streams. Furthermore, the particles in each of those high density streams flow in the same direction thus providing further evidence that we are witnessing a streaming instability. We also see that the densest clumps in Figure 10b seem to form in regions where multiple streams intersect. We pick out the densest such clump from Figure 10b (shown with a box around it) and magnify that 12x12 zone region in Figure 10c. Figure 10c shows the positions of the particles with colored lines coming out of them that correspond to the directions and magnitudes of their velocities. The length of the line coming out of each particle is proportional to the magnitude of its velocity. The color of the line coming out of each particle is proportional to the direction of its velocity in the xy-plane. Figure 10c clearly shows us that the densest clump is made of two streams that intersect at the point of maximal density. As a result we have conclusively shown the role that multiple, intersecting streams in the streaming instability play in forming the densest clumps.

It is also very interesting to visually demonstrate that the RMS velocity of the dust within a particular stream is much reduced relative to the mean velocities of the dust between streams. To illustrate that, in Figure 11 we focus on the same 12x12 zone region from Figure 10b. However, this time we have selected only those particles whose velocity in the x-y plane that lies in a narrow range of angles, between -135° and 135° (as shown in the colour-coded pie chart at the bottom of Figure 11). The range of angles has, however, to be selected to match the velocity of the dominant stream that one is trying to select. The velocity of the particles relative to the mean streaming velocity is also shown as colored lines coming out of the particles. The sequence of numbers that are overlaid on Figure 11 show the local RMS velocity in meters per second of the particles at that point relative to the mean streaming velocity. Figure 11 is very illustrative because it clearly shows that the RMS velocity of the particles relative to the local streaming velocity is reduced substantially in the vicinity of the densest clumps. In the next subsection we discuss the implications of the streaming instability on grain collisions and growth.

## 4.5) Implications of the streaming instability for grain collisions and growth

The process of dust grain growth is sensitive to the relative velocities between grains that are to stick. We assess this issue by interpolating the particles to the mesh and calculating the RMS velocity dispersion of dust grains in each zone. Recall that we plot the velocity dispersion of the dust grains of a given family with the density of dust in the same zone in Figure 9. The regions with the highest dust density have a small velocity dispersion of less than 0.01 $c_s$. In our 10 au simulations the sound speed is 0.68 km s$^{-1}$, and so the densest regions of dust have velocity dispersions less than 6.8 m s$^{-1}$. The experimental results of Blum & Wurm (2000) indicate that collisions of dust particles will result in fragmentation rather than growth if the relative velocities are larger than a threshold velocity given by $v_{\text{thresh}} = 1.9$ m s$^{-1}$. We find that most of the dust of any size has velocity dispersions larger than this threshold velocity, but 15.6% of the mass of 2.5 mm dust has a velocity dispersion less than 3 $v_{\text{thresh}}$, and 4% of the 2.5 mm dust grains have a velocity dispersion less than 2 $v_{\text{thresh}}$ (only 0.03% of the dust is found in zones with a velocity dispersion less than $v_{\text{thresh}}$). We see therefore that some of the dust in our simulations is capable of undergoing grain growth. As one goes to larger radii, we expect the streaming instability to be more vigorous which makes the conditions even more favorable for grain growth. This tendency for grain growth will therefore be examined at several radial stations for a minimum mass solar nebula in the sequel paper. In the present paper we focus more deeply on the collision times and mean free path of the dust at 10 au. Previous studies of grain coagulation and growth, see for example Dullemond & Dominik (2005), have had to focus on phenomenological estimates of grain growth with no attention paid to the dynamics of grains as they respond to accretion disc turbulence. It is therefore worth pointing out that simulations like the ones presented here could serve as a template of a more dynamically-oriented study of grain coagulation and growth.

The cross-section $\sigma$ of an uncharged, spherical dust grain of the type simulated here is proportional to the square of its radius. As a result, for dust particles that have a physical number density given by $n_d$ we have a mean free path given by $\lambda_d = 1/(\sigma\, n_d)$.

We therefore expect the largest dust grains that undergo streaming instability to have the smallest mean free paths in the regions where their density is the largest. If the dust particles have an RMS velocity $v_{d,rms}$ between themselves then the collision time is given by $\tau_c = \lambda_d / v_{d,rms}$. Thus we expect the collision times to be extremely short in the clumps. Figure 12a shows a color plot of the mean free path as a ratio of the average mean free path in the midplane for the largest dust in the 10 au simulation. We see clearly that the mean free path in the clumps is two orders of magnitude smaller than the average mean free path in that plane. Figure 12b shows a color plot of the collision time as a ratio of the average collision time again shown in the midplane for the largest dust in the 10 au simulation. We see that the collision time for the dust in the clumps is two orders of magnitude smaller than the average collision time. It is also worth pointing out that Figures 12a and 12b only pertain to the collision of one 2.5 mm grain against another similar grain. This is of course the simplest thing that one can display. It should be remembered though that the 250 μm grains also participate in the streaming instability and their clumps directly overlie those of the 2.5 mm grains. As a result, if a favorable mechanism is found to make smaller grains adhere to the larger grains, the tendency for grain growth would be significantly enhanced because of the larger number density of the smaller grains.

An enhancement in the densities of the large dust has important implications for the growth of dust grains. Dust with sizes in the range of 10 cm – 1 m are large enough to see significant drag from the sub-Keplerian gas disc, while not large enough that their inertia can minimize this drag force. As a result, grains in this range will lose angular momentum quickly, and can accrete inwards to the protostar in only a few orbital timescales. In order to form the boulder-sized objects that are needed in the standard model of planet formation, growth through the decimeter-size regime must be fast. Our results indicate that these large dust grains are likely to be highly clumped. As collisional processes proceed at a rate proportional to density squared, the clumping in the disc midplane is a favorable scenario for the rapid growth of boulders from dust.

**5) Conclusions**

In this paper we have used the RIEMANN code for computational astrophysics to study the interaction of an MRN distribution of dust grains with gas at specific radial locations in a vertically stratified protostellar accretion disc. The disc was modeled to have the density and temperature of a minimum mass solar nebula and shearing box simulations at radii of 0.3 and 10 au are reported here. The disc was driven to a fully-developed turbulence via the MRI. After verifying that our simulation code accurately collective instabilities in the dust, we have found the following interesting results:

1) We have shown that for normal dust to gas ratios the sedimentation of dust to the midplane of an accretion disc has very little effect on the MRI-driven turbulence.

2) We have shown that the dust to gas scale heights follow a very simple scaling law where $H_{dust}/H_{gas} \propto a^{-1}$ with "a" being the size of the dust. This relation is true for small, spherical dust grains at the two radii studied here. The RMS velocity of the dust also scales as the turbulent velocity. As a result, the results for dust scale heights are consistent with the theory of Dubrulle, Morfill & Sterzik (1995). The theory can therefore be used quite successfully in explaining observations of dust to gas ratios is protostellar discs as was done in Rettig et al. (2006).

3) The larger dust grains that sediment to the midplane of the accretion disc in our 10 au simulation show a pronounced tendency to form clumps. We identify the tendency to form clumps with the streaming instability of Youdin & Goodman (2005).

4) We show that the RMS velocity in the regions with the highest density of large dust grains is lower than the mean RMS velocity in the grains. We also find that the streams that give rise to the streaming instability have their velocities aligned in the same direction. The streaming instability is seen to be very vigorous and persistent once it forms. The densest clumpings of large dust are shown to form where the streams intersect.

5) We examined the implications of this reduced RMS velocity in the densest streams for dust collision and growth. We find that the mean free path for collisions between dust particles as well as the collision time is reduced by almost two orders of magnitude relative to the average mean free path and collision time.

6) Blum and Wurm (2000) showed that dust collisions only result in sticking if the relative velocities between colliding grains is less than a certain threshold value. We show that some of the dust in our 10 au simulations indeed has relative velocities that fall below this threshold, making if favorable to have some grain growth.

7) Our models show that larger dust can clump to produce densities that are almost two orders of magnitude higher than the mean dust density. Models of coagulation and growth posit that the growth rate of dust is proportional to the square of the dust density, see for example Dullemond & Dominick (2005). Even if we apply those simple models for grain growth to the clumps, we realize that the growth rate of dust in the clumps will be dramatically enhanced relative to the mean growth rate.

8) Our simulations provide a dynamically consistent picture for the distribution of dust which could potentially be very useful for those who model SEDs of discs as well as for those who model dust coagulation and growth.

**References**

Balbus, S. A. & Hawley, J. H., 1991, ApJ, 376, 214

Balbus, S. A. & Hawley, J. H., 1998, Rev. Mod. Phys., 71, 1

Balsara, D. S., 1998a, ApJS, 116, 119

Balsara, D. S., 1998b, ApJS, 116, 133

Balsara, D. S., 2004, ApJS, 151, 149

Balsara, D. S., & Spicer, D., 1999a, J. Comput. Phys, 148, 133

Balsara, D. S., & Spicer, D., 1999b, J. Comput. Phys, 149, 270

Blum, J. & Wurm, G., 2000, Icarus, 143, 138

Boss, A., 2005, ApJ, 629, 535

Bouwman, J., de Koter, A., Dominik, C., Waters, L. B. F. M., 2003, A&A, 401, 577

Brearley, A.J. & Jones, R.H., 1998, Reviews in Mineralogy, 36, Planetary Materials, ed
    J.J. Papike (Washington DC : Mineralogical Society of America)

Burrows, C. J. et al., 1996, ApJ, 473, 437

Carballido, A., Fromang, S. & Papaloizou, J., 2006, MNRAS, 373, 1633

Chiang, E. I. & Goldreich, P., 1997, ApJ, 490, 368

Cotera, A. S. et al., 2001, ApJ, 556, 958

Cuzzi, J. N., Dobrovolskis, A. R., & Champney, J. M., 1993, Icarus, 106, 102

D'Alessio, P., Canto, J., Calvet, N., & Lizano, S., 1998, ApJ, 500, 411

D'Alessio, P., Calvet, N., Hartmann, L., Franco-Hernández, R., & Servín, H., 2006, ApJ,
    638, 314

Dubrulle, B., Morfill, G., & Sterzik, M., 1995, Icarus, 114, 237

Dullemond, C. P. & Dominik, C., 2005, A&A, 434, 971

Fatuzzo, M., Adams, F. C., Melia, F., 2006, ApJ, 653, 49

Fromang, S. & Papaloizou, J., 2006, A&A, 452, 751

Fromang, S. & Papaloizou, J., 2007, A&A, 476, 1113

Fromang, S., Papaloizou, J., Lesur, G., & Heinemann, T., 2007, A&A, 476, 1123

Gammie, C. F., 1996, 457, 355

Garaud, P. Barriere-Fouchet, L. & Lin, D.N.C., 2004, ApJ, 603, 292

Garaud, P. & Lin, D.N.C., 2004, ApJ, 608, 1050

Goldreich, P., Lithwick, Y. & Sari, R., 2004a, ARA&A

Goldreich, P., Lithwick, Y. & Sari, R., 2004b, ApJ, 614, 497

Goldreich, P. & Ward, W.R., 1973, ApJ, 183, 1051

Goldreich, P. & Lynden-Bell, D., 1965, MNRAS, 130, 125

Goldreich, P. & Tremaine, S., 1978, ApJ, 222, 850

Gomez, G.C. & Ostriker, E.C., 2005, ApJ, 630, 1093

Goodman, J. & Pindor, B., 2000, Icarus, 148, 537

Hartmann, L. "Accretion Processes in Star Formation", New York: Cambridge UP, 1998

Ishitsu, N. & Sekiya, M., 2003, Icarus, 165, 181

Johansen, A., Henning, T. & Klahr, H., 2006, ApJ, 643, 1219

Johansen, A. & Youdin, A., 2007, ApJ, 662, 627

Mathis, J. S., Rumpl, W., Nordsieck, K. H. 1977, ApJ, 217, 425

Miller, K. A., & Stone, J. M., 2000, ApJ 534, 398

Miyake, K. & Nakagawa, Y., 1995, ApJ, 441, 361

Mizuno, H., 1980, Progress in Theoretical Physics, 64, 544

Nakagawa, Y., Sekiya, M. & Hayashi, C., 1986, Icarus, 115, 304

Natta, A., Testi, L., Calvet, N., Henning, Th., Waters, R., Wilner, D. "Dust in Protoplanetary Disks: Properties and Evolution", Protostars and Planets V, 2007, ed. B. Reipurth, D. Jewitt & K. Keil, p. 767

Pollack, J. B., Hubickyj, O., Bodenheimer, P., Lissauer, J. J., Podalak, M., & Greenzweig, Y., 1996, Icarus, 124, 62

Rettig, T., Brittain, S., Simon, T., Gibb, E., Balsara, D. S., Tilley, D. A., Kulesa, C., 2006, ApJ, 646, 342

Rodmann, J, Henning, T., Chandler, C.J., Mundy, L.G. & Wilner, D.J., 2006, A&A, 336, 221

Safronov, V.S. ,1969, Evolutsiia doplanetnogo oblaka. (Moscow: Nakua)

Sekiya, M., 1998, Icarus, 133, 298

Sicilia-Aguilar A., Hartmann L.W., Hernandez J., Briceo C., Calvet N., 2005, AJ, 130, 188

Suttner, G. , & Yorke, H., 2001, ApJ, 551, 461

Tilley, D. A., & Balsara, D. S., 2008, MNRAS, 389, 1058

Throop, H.B. & Bally, J. 2005, ApJ, 623, L149

Umurhan, O. M., Menou, K., Regev, O., 2007, Phys. Rev. L., 98, 4501

Umurhan, O.M., Regev, O. & Menou, K., 2007, Phys. Rev. E, 76, 6310

Van Boekel et al., 2004, Nature, 432, 479

Weidenschilling, S. J., 1977, MNRAS, 180, 57

Weidenschilling, S. J., 1980, Icarus, 44, 172

Weidenschilling, S. J., 2006, Icarus, 181, 572

Yan, H. & Lazarian, A., 2003, ApJ, 592, L33

Yorke, H.W., 1979, A&A, 80, 308

Youdin, A.N. & Chiang, E.I., 2004, ApJ, 601, 1109

Youdin, A.N. & Goodman, J., 2005, ApJ, 620, 459

Youdin, A.N. & Johansen, A., 2007, ApJ, 662, 613

Youdin, A.N. & Shu, F.H., 2002, ApJ, 580, 494

Table 1 : Physical parameters for the various simulations.

| Run Name | r (AU) | $\Omega$ (sec$^{-1}$) | $\rho_0$ (gm/cm$^3$) | $T_0$ (K) | $c_s$ (cm/sec) | $H_{gas}$ (AU) | $\beta_0$ | $N_x,N_y,N_z$ |
|---|---|---|---|---|---|---|---|---|
| S0.3 | 0.3 | 8.568 x 10$^{-7}$ | 2.666 x 10$^{-8}$ | 365.0 | 1.326 x 10$^5$ | 0.01553 | 400 | 96,96,192 |
| S10 | 10 | 4.452 x 10$^{-9}$ | 1.405 x 10$^{-12}$ | 95.85 | 6.797 x 10$^4$ | 1.532 | 400 | 96,96,192 |

Table 2 : Ratio of particle stopping time to dynamical time ($t_s \Omega$) for various radii and various dust sizes in a minimum mass solar nebula. The Stokes, Epstein and Turbulent regimes are denoted by S, E and T respectively.

| dust size /radius | 0.3AU | 10 AU |
|---|---|---|
| 0.1 μ | 4.847e-9 E | 9.324e-7 E |
| 1 μ | 4.847e-8 E | 9.324e-6 E |
| 10 μ | 4.847e-7 E | 9.324e-5 E |
| 100 μ | 4.847e-6 E | 9.324e-4 E |
| 1 mm | 5.905e-5 S | 9.324e-3 E |
| 1 cm | 5.905e-3 T | 9.324e-2 E |
| 10 cm | 5.905e-1 T | 9.324e-1 E |

Table 3 : Radial drift velocities (in m/s) for dust of various sizes at various radii in a minimum mass solar nebula

| dust size / radius | 0.3AU | 10 AU |
|---|---|---|
| 0.1 μ | 7.436e-5 | 2.090e-2 |
| 1 μ | 7.436e-4 | 0.2090 |
| 10 μ | 7.436e-3 | 2.090 |
| 100 μ | 7.436e-2 | 20.90 |
| 1 mm | 0.9059 | 208.9 |
| 1 cm | 90.59 | 2.071e3 |
| 10 cm | 6.751e3 | 1.128e4 |

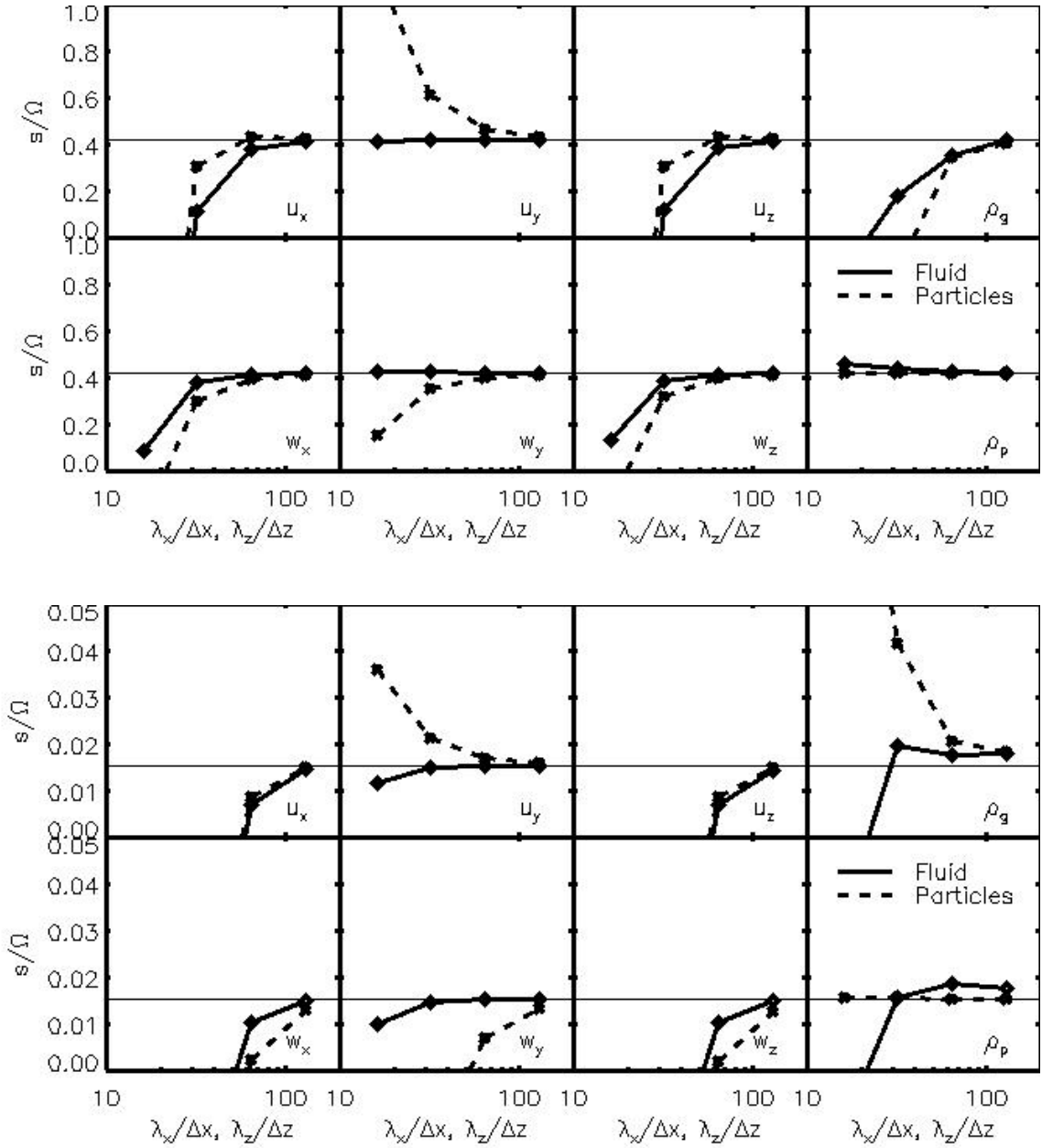

Figure 1: Growth rates of the streaming instability in an unstratified disc. (Top) Growth rates of gas velocity, gas density, dust velocity, and dust density for model 'linA'. (Bottom) Growth rates for model 'linB'.

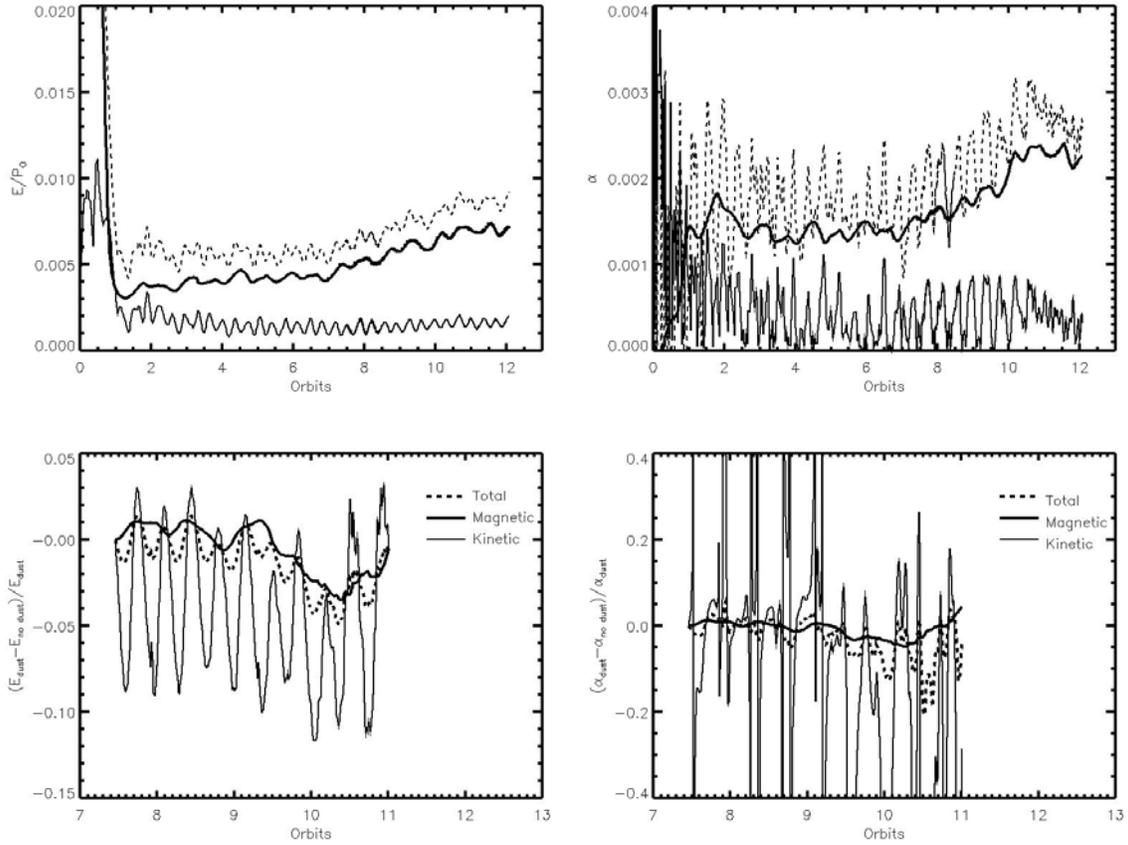

Figure 2. (a) Evolution of the magnetic energy (thick solid line), kinetic energy (thin solid line), and total (kinetic + magnetic) energy (dashed line). (b) Evolution of the magnetic stress (thick solid line), kinetic stress (thin solid line), and total stress (dashed line). (c) Difference between the magnetic, kinetic, and total energies between a simulation with dust and an equivalent simulation without dust. (d) Difference between the magnetic, kinetic, and total stresses between a simulation with dust and an equivalent simulation without dust.

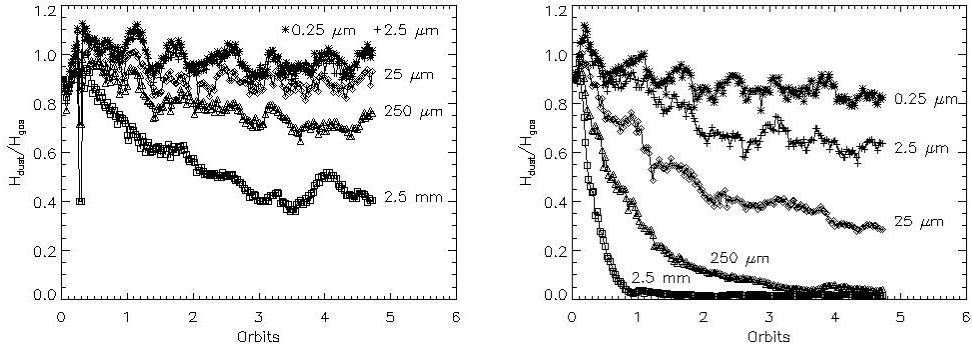

Figure 3. Scale height of the dust of different grain sizes as a function of time, at 0.3 au (a) and 10 au (b). Stars indicate 0.25 µm dust, crosses indicate 2.5 µm dust, diamonds indicate 25 µm dust, triangles indicate 250 µm dust, and squares indicate 2.5 mm dust. At 0.3 au, the 0.25 µm and 2.5 µm dust nearly overlap.

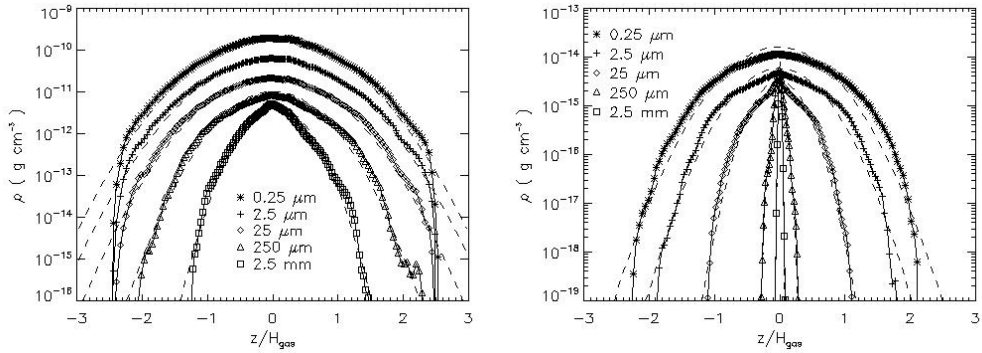

Figure 4. Mean dust density as a function of height above and below the midplane. (a) Dust density at 0.3 au (b) Dust density at 10 au  Dashed lines are fits of a Gaussian profile to the data.

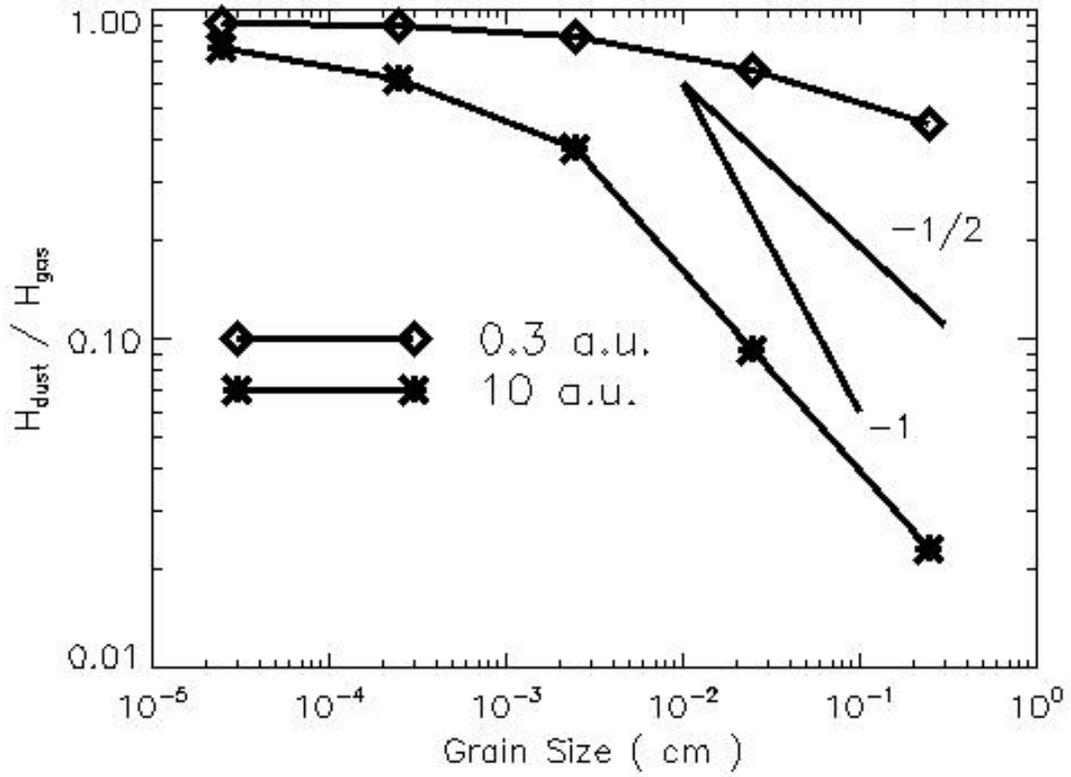

Figure 5. Scale height of the dust of various sizes as a function of the size of the dust grains, after the dust has reached a steady state.

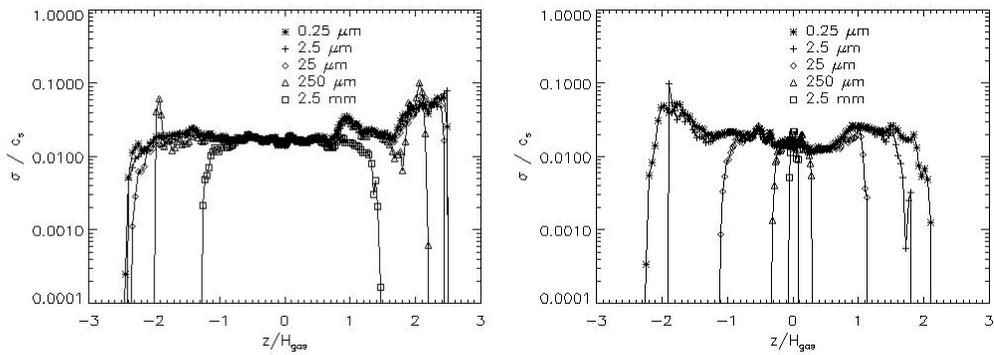

Figure 6. Velocity dispersion of the dust as a function of height above and below the midplane. (a) Velocity dispersion at 0.3 au (b) Velocity dispersion at 10 au

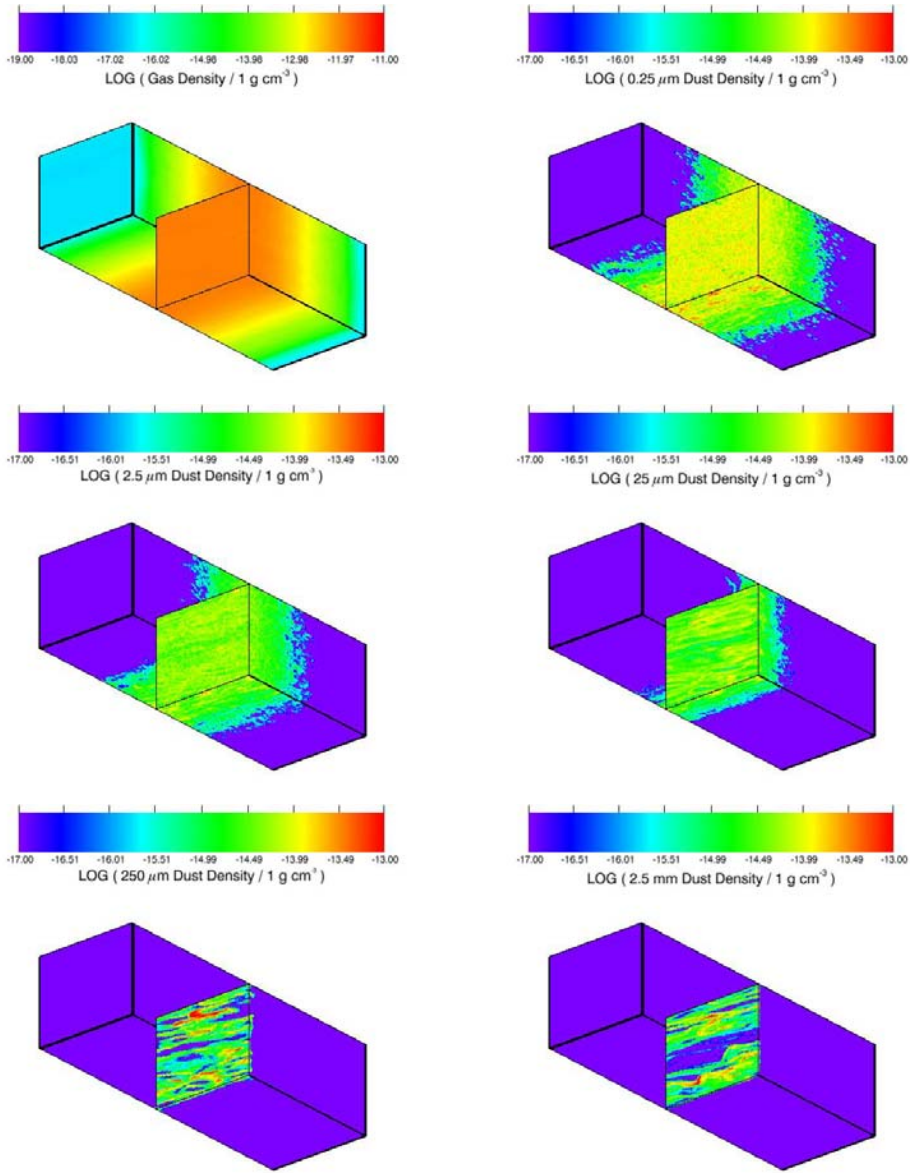

Figure 7. Images showing the dust density in four planes within the simulation: midplane (z=0, in the centre of each image), top plane (z = 3 $H_{gas}$, on the top-left of each image), outer plane (x = $H_{gas}$, on the bottom of each image), and rear plane (y = $H_{gas}$, on the upper right of each image). The upper two frames show the gas density and the density of the 0.25 µm dust. The middle two frames show the 2.5 µm dust density and the 25 µm dust density. The lowest two frames show the 250 µm dust density and the 2.5 mm dust density.

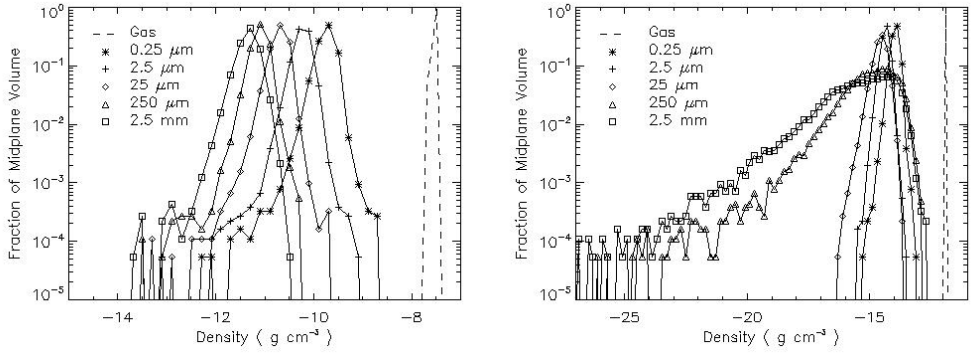

Figure 8. Histograms of the dust density in the midplane of the disc, showing the fraction of the volume between z=-0.014 $H_{gas}$ and z=0.014 $H_{gas}$ that contains dust of that density. (a) Histogram of dust density at 0.3 au  (b) Histogram of dust density at 10 au

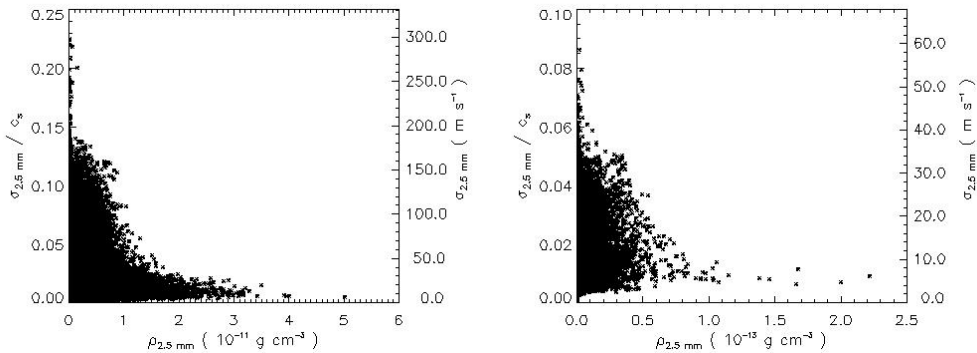

Figure 9. Velocity dispersion in each zone for the 2.5 mm dust versus the density in that zone. (a) 0.3 au  (b) 10 au

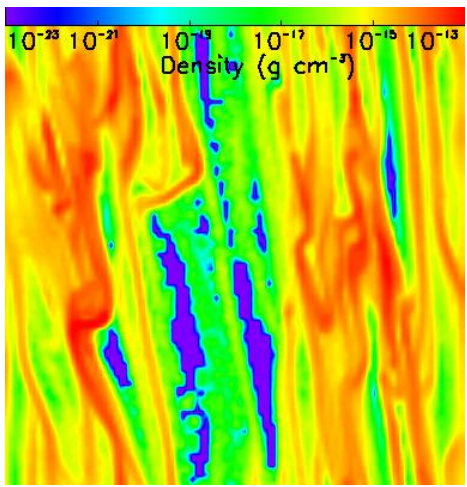

Figure 10. (a) Density of 2.5 mm dust in the midplane.

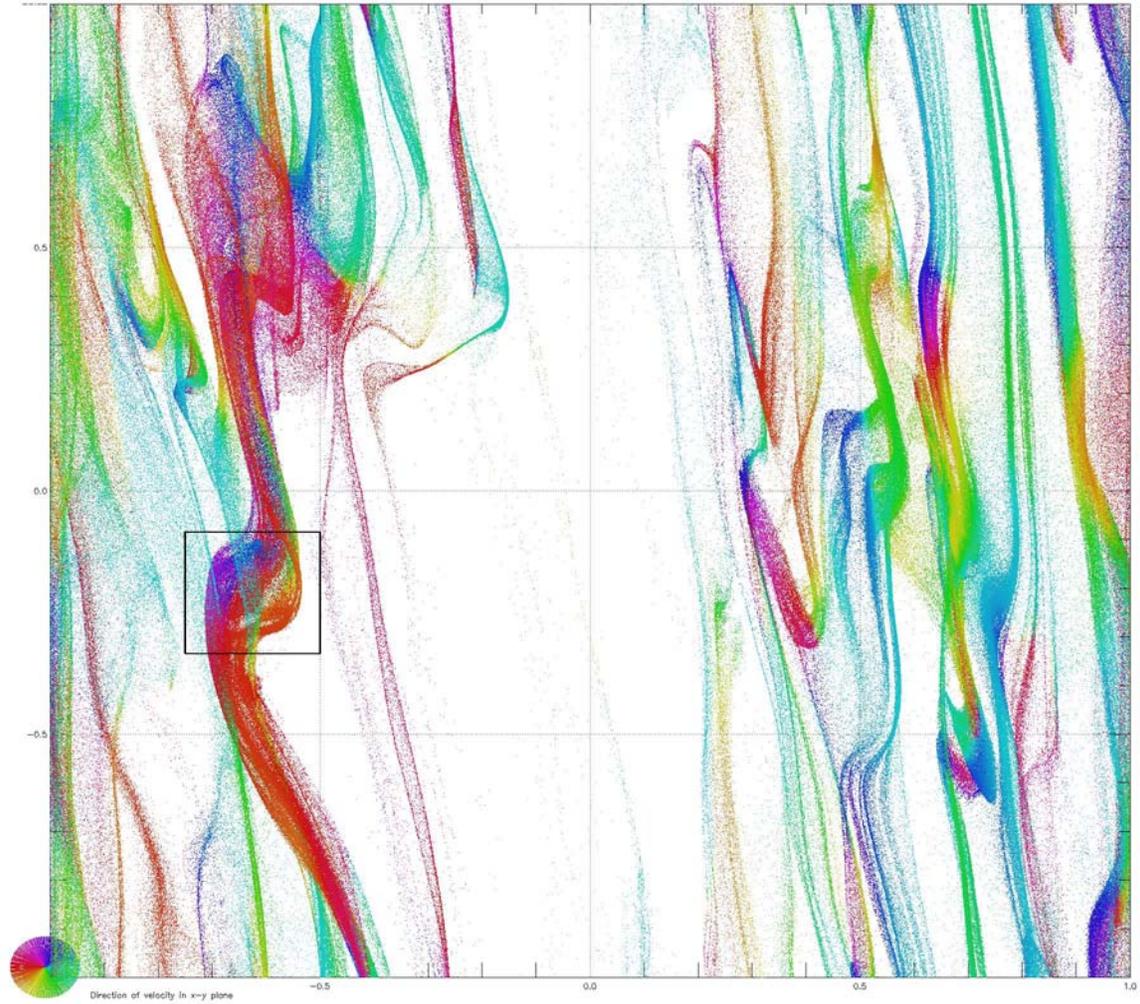

Figure 10 (cont.) (b) Projection of the locations of the 2.5 mm dust superparticles, viewed from above the midplane. The colour coding indicates the direction of each superparticle's velocity in the midplane, after the shear has been removed, as indicated by the coloured pie chart in the lower left-hand corner. The box indicates the region examined in Figure 10c.

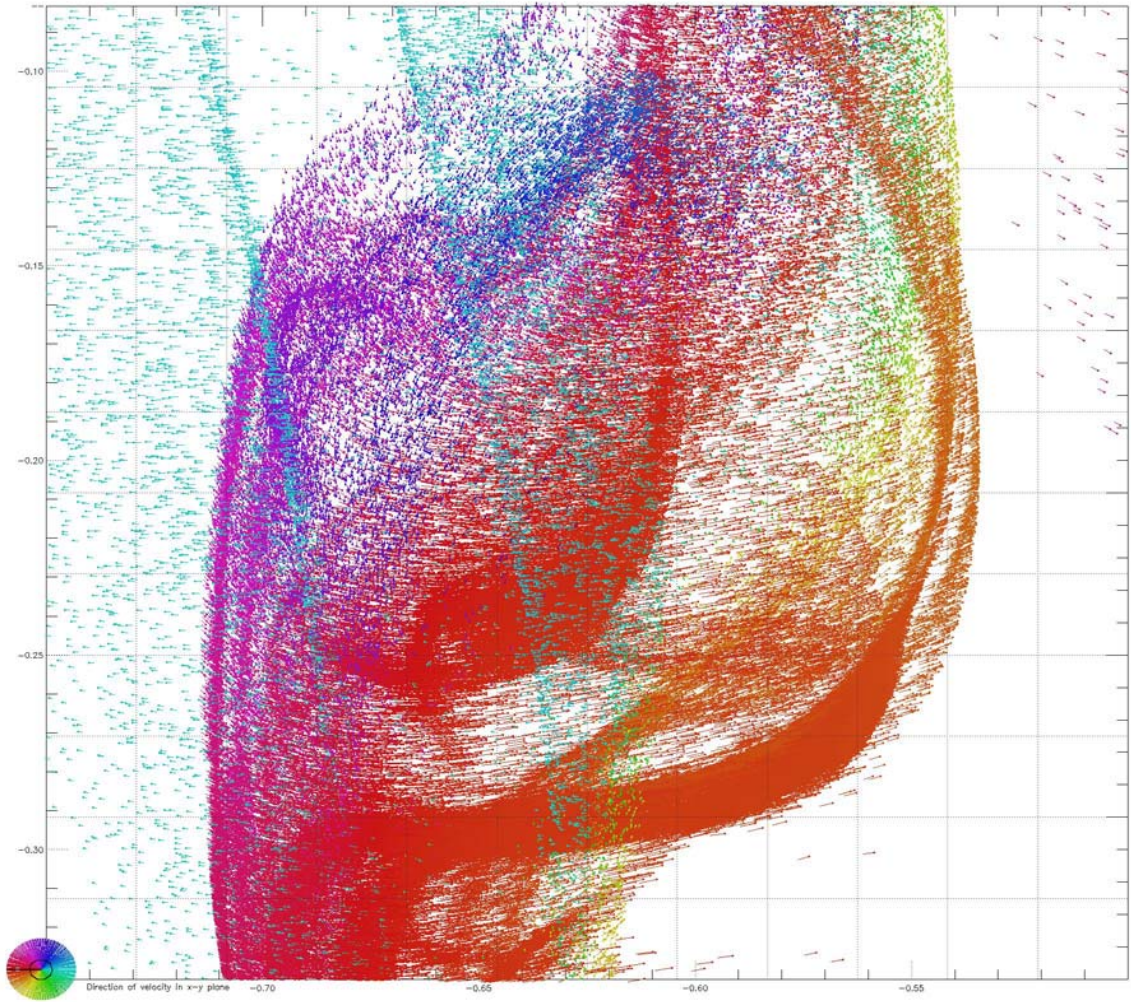

Figure 10 (cont.) (c) A view of the region marked by the box in Figure 10b, showing the positions of the superparticles of 2.5 mm dust as viewed from above the midplane. The velocity in the x-y plane of each particle is marked both through the colour of the particle, according to the pie chart in the lower left-hand corner, as well as by a vector.

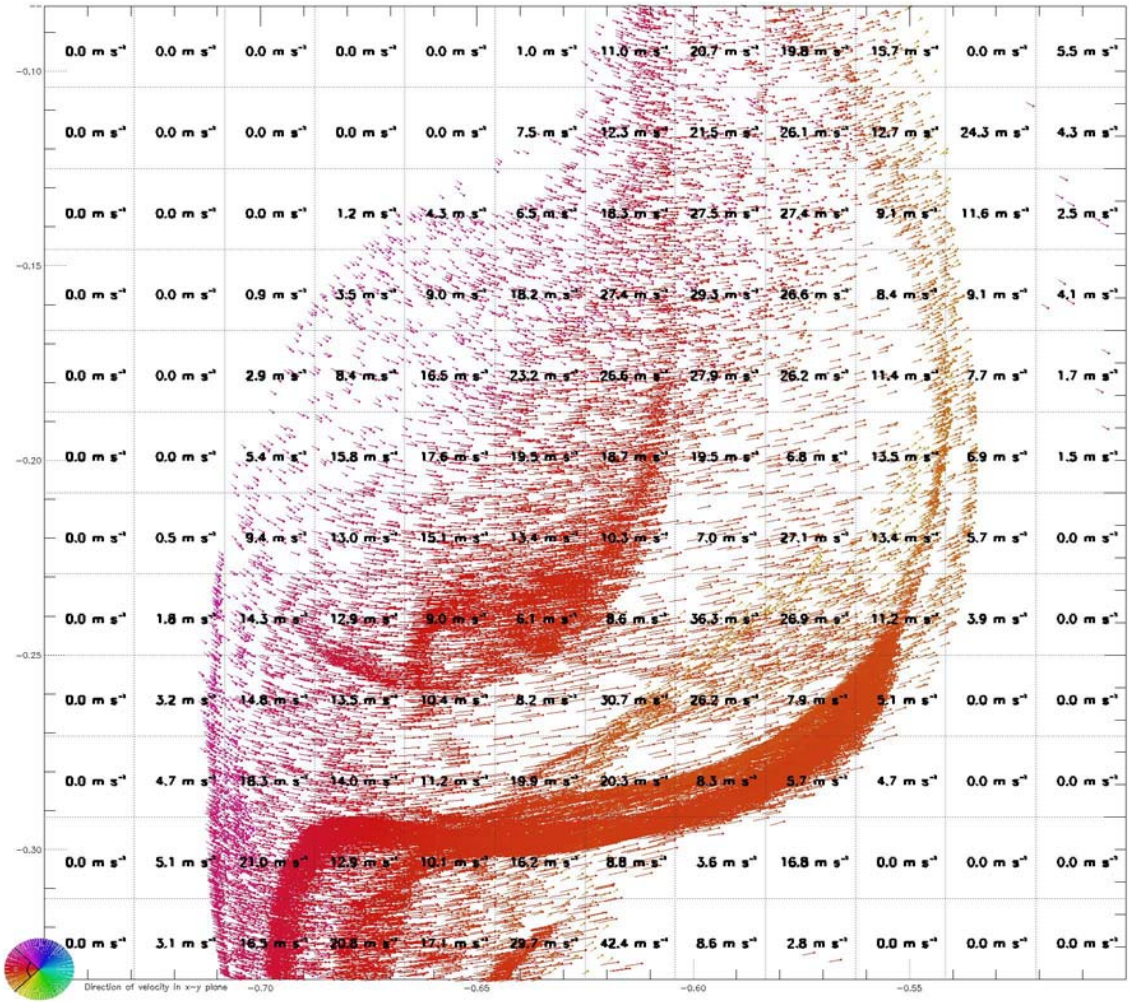

Figure 11. The same view as in Figure 10c, but only showing the superparticles that have velocity vectors (after the shear in the disc has been removed) that are within 45° of 180°, as indicated in the colour-coded pie chart in the lower left-hand corner of the plot. This isolates just a few of the streams that are in this region of the domain. For each grid zone, we label the velocity dispersion of this subset of superparticles. For clarity, we plot only one out of every four superparticles.

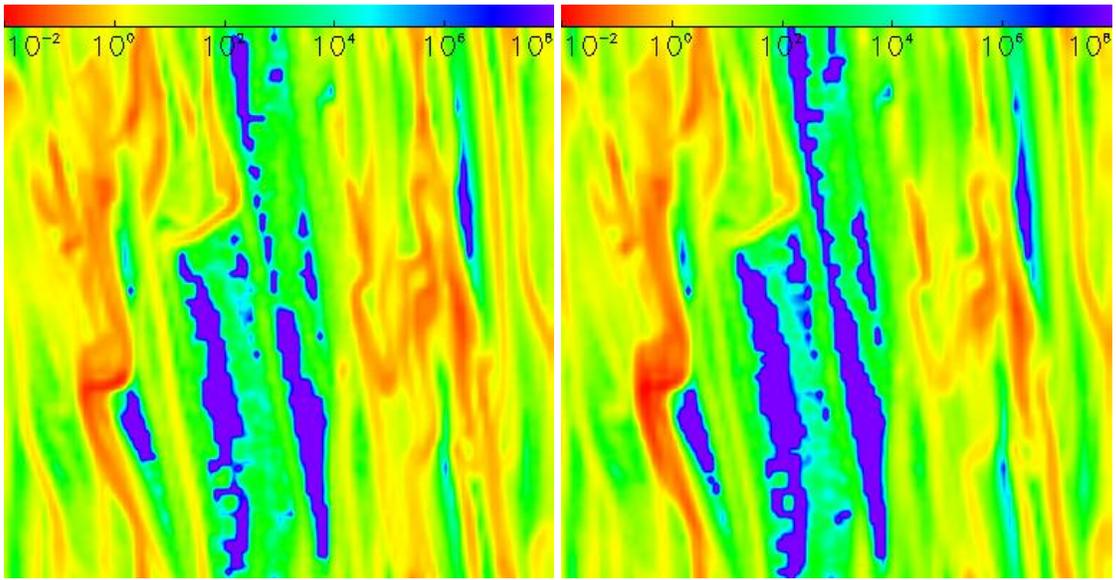

Figure 12. (a) Ratio of the mean free path of 2.5 mm dust grains in each zone to the average mean free path in the midplane. (b) Ratio of the collision time of 2.5 mm dust grains in each zone to the average collision time in the midplane.